\documentclass[sigconf,review=false,anonymous=false]{acmart}
\acmConference[ICSE 2024]{46th International Conference on Software Engineering}{April 2024}{Lisbon, Portugal}

\usepackage{xspace}
\usepackage{booktabs}
\usepackage{paralist}
\usepackage{tcolorbox}
\usepackage{wrapfig}
\usepackage{tikz}
\usepackage{pgfplots}
\usepackage{blindtext}
\usepackage{mdframed}

\newcommand{\RQ}[1]{\texorpdfstring{RQ\textsubscript{#1}}{RQ#1}}

\newcommand{\approach}[0]{\textsc{Frenilla}\xspace}
\newcommand{\scissor}[0]{\textsc{SDC-Scissor}\xspace}
\newcommand{\beamng}[0]{\textsc{BeamNG.tech}\xspace}
\newcommand{\frenetic}[0]{\textsc{Frenetic}\xspace}
\newcommand{\freneticV}[0]{\textsc{FreneticV}\xspace}

\title{Diversity-guided Search Exploration for Self-driving Cars Test Generation through Frenet Space Encoding}

\author{Timo Blattner}
\email{timo.blattner@students.unibe.ch}
\affiliation{
  \institution{University of Bern}
  \country{Switzerland} 
}

\author{Christian Birchler}
\email{christian.birchler@{zhaw,unibe}.ch}
\affiliation{
  \institution{Zurich University of Applied Sciences \\ University of Bern}
  \country{Switzerland}
}

\author{Timo Kehrer}
\email{timo.kehrer@unibe.ch}
\affiliation{
  \institution{University of Bern}
  \country{Switzerland}
}

\author{Sebastiano Panichella}
\email{sebastiano.panichella@zhaw.ch}
\affiliation{
  \institution{Zurich University of Applied Sciences}
  \country{Switzerland}
}

\begin{abstract}
The rise of self-driving cars (SDCs) presents important safety challenges to address in dynamic environments.
While field testing is essential, current methods lack diversity in assessing critical SDC scenarios.
Prior research introduced simulation-based testing for SDCs, with Frenetic, a test generation approach based on Frenet space encoding, achieving a relatively high percentage of valid tests (approximately 50\%) characterized by naturally smooth curves.
The ``minimal out-of-bound distance'' is often taken as a fitness function, which we argue to be a sub-optimal metric.
Instead, we show that the likelihood of leading to an out-of-bound condition can be learned by the deep-learning vanilla transformer model.
We combine this ``inherently learned metric'' with a genetic algorithm, which has been shown to produce a high diversity of tests.
To validate our approach, we conducted a large-scale empirical evaluation on a dataset comprising over 1,174 simulated test cases created to challenge the SDCs behavior.
Our investigation revealed that our approach demonstrates a substantial reduction in generating non-valid test cases, increased diversity, and high accuracy in identifying safety violations during SDC test execution.
\end{abstract}

\begin{document}
\maketitle

\section{Introduction}
The increasing autonomy and widespread adoption of autonomous systems \cite{SBFT-UAV2024,khatiri2023simulation} such as self-driving cars (SDCs) pose significant challenges regarding their operational safety in dynamic environments \cite{us-adas-accidents,NPR,10190377,UAVtosem}. 
While field testing is crucial to assess the behavior of autonomous systems in different conditions, there is still a big gap in testing SDC features in diverse and critical scenarios that faithfully replicate the complexity of real-world dynamics \cite{KhatiriPT23,10.1145/3533818}.

Previous work proposed simulation-based testing environments in the context of SDCs to assess the lane-keeping ability of the vehicle, with important results in efficiently identifying test cases that effectively reveal faults of SDCs in different environmental conditions \cite{10190377,DBLP:journals/ese/BirchlerKBGP23,9825849,PanichellaGZR21}.
Notably, the Frenetic test generation approach \cite{KlikovitsCCA23,WinstenP23}, based on Frenet space encoding, emerged as one of the most successful approaches for generating a high percentage of valid test cases (approximately 50\%) with naturally smooth curves.
However, Frenetic still has two important limitations: (i) it generates a high percentage of invalid test cases (e.g., excessive curvature) \cite{KlikovitsCCA23,WinstenP23}; and (ii) it relies on the "out of bound distance", the distance between the center of the center line of the road, which we argue to be a sub-optimal fitness function: good drivers will always push to the outside first before turning in to reduce the centripetal acceleration and extend the corner. Consequently, the metric might show high fitness for a test where the driver actually drives correctly. 


To address this gap, we propose an approach named \approach, which employs a transformer model that has been trained to predict the likelihood of an out-of-bound condition at each point on the road.
We combine this discriminator with a genetic algorithm to guide the search exploration toward more diverse and valid test cases while keeping a competitive fault detection ability.
To validate our approach, we conducted a large-scale empirical study on a dataset comprising over 1,174 simulated test cases created to challenge the SDC's behavior, which allowed us to explore the relation between road shape compositions and SDC safety violations.


\section{Background and related work}
The \approach approach works based on the concept of Frenet frames. 
Hence, we overview that model along with how \frenetic was using it in previous work.

The road shape of an SDC test case can be modeled as a series of curves $s(t)$.
Furthermore, certain mathematical properties can be expressed on curves.
For instance, the \textit{Frenet-Serret formulas} describe the dynamics of a moving object among a curve.
Specifically, $(T, N, B)$ is a \textit{frenet frame}, where $T$ is the tangent on the curve heading the object's movement, $N$ is the normal unit vector of $T$, and $B$ is a unit vector by the cross product of $T$ and $N$.

Moreover, $\frac{d}{ds}$ is the derivative with respect to arclength, $\kappa$ is the curvature, and $\tau$ is the torsion of the curve.
The two scalars $\kappa$ and $\tau$ effectively define the curvature and torsion of a space curve.
The associated collection, T, N, B, $\kappa$, and $\tau$, is called the Frenet–Serret apparatus.
Intuitively, curvature measures the failure of a curve to be a straight line, while torsion measures the failure of a curve to be planar.



Castellano et al.~\cite{DBLP:conf/sbst/CastellanoCTKZA21} make use of the Frenet representation in their genetic algorithm, which was shown to be one of the most successful approaches at the SBFT 2021 competition.
Peltomäki et al.~\cite{peltomaki2022wasserstein} also use the Frenet representation with a Wasserstein Generative Adversarial Network to estimate the fitness of tests before having to execute them on the simulator.
Intuitively, they generate tests with higher fitnesses and the discriminator gets better at identifying tests with lower fitnesses.
However, they limit themselves to segments with 6 points and struggle with generating valid tests. 

\section{The \approach approach} 

While we utilize the dynamic encoding technique of Frenetic, we tackle the abovementioned limitations of Frenetic in this section. Specifically, in terms of novelty, our approach integrates a transformer model that implicitly learns the out-of-bound condition and expected outcome, supposed to improve the overall robustness and effectiveness of state-of-the-art approaches.

\subsection{Seed Data}
\begin{figure}[t]
    \centering
    \includegraphics[width=0.5\linewidth]{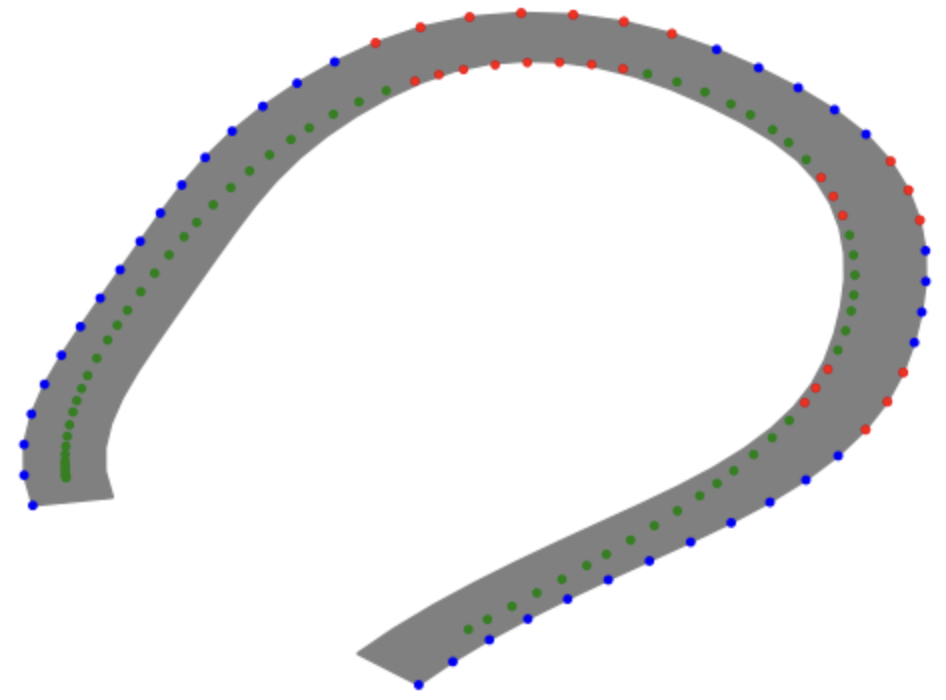}
    \caption{The road in grey, with the simulation trace in green showing the out-of-bounds conditions in red}
    \label{fig:oob-trace}
    \vspace{-5mm}
\end{figure}

Given several failing simulation tests, we know where the car went out of bounds (Figure~\ref{fig:oob-trace}). We match those points to the closest curvature point in our road segment, which we label as out-of-bound.
So, we have a pairing of curvature points, and if they lead to an out-of-bound condition, we can use them for training.

We re-used previously in simulation-executed SDC test cases consisting of information regarding the SDC's trajectory. In total, \approach is trained on 12,561 road segments having a length of 50 meters. The seed data is also available in the replication package of~\cite{DBLP:journals/ese/BirchlerKBGP23}.


\subsection{Discriminator}

In our framework, each road is represented by a series of vectors $(c_i, s_i)$, where $c_i$ indicates the curvature and $s_i$ the cumulative step size.
Each vector is encoded into a 128-dimensional vector, which is fed to a vanilla transformer with 6 heads and 6 layers, with a linear layer at the end predicting the in and out of bounds.
The model is trained with a block size of 50 (representing the length of a road), batch size of 1024, and Adam optimizer with a learning rate of 3e-4 with a dropout of 0.2 in the linear layers.

\begin{figure}[t]
    \centering
    \includegraphics[width=0.9\linewidth]{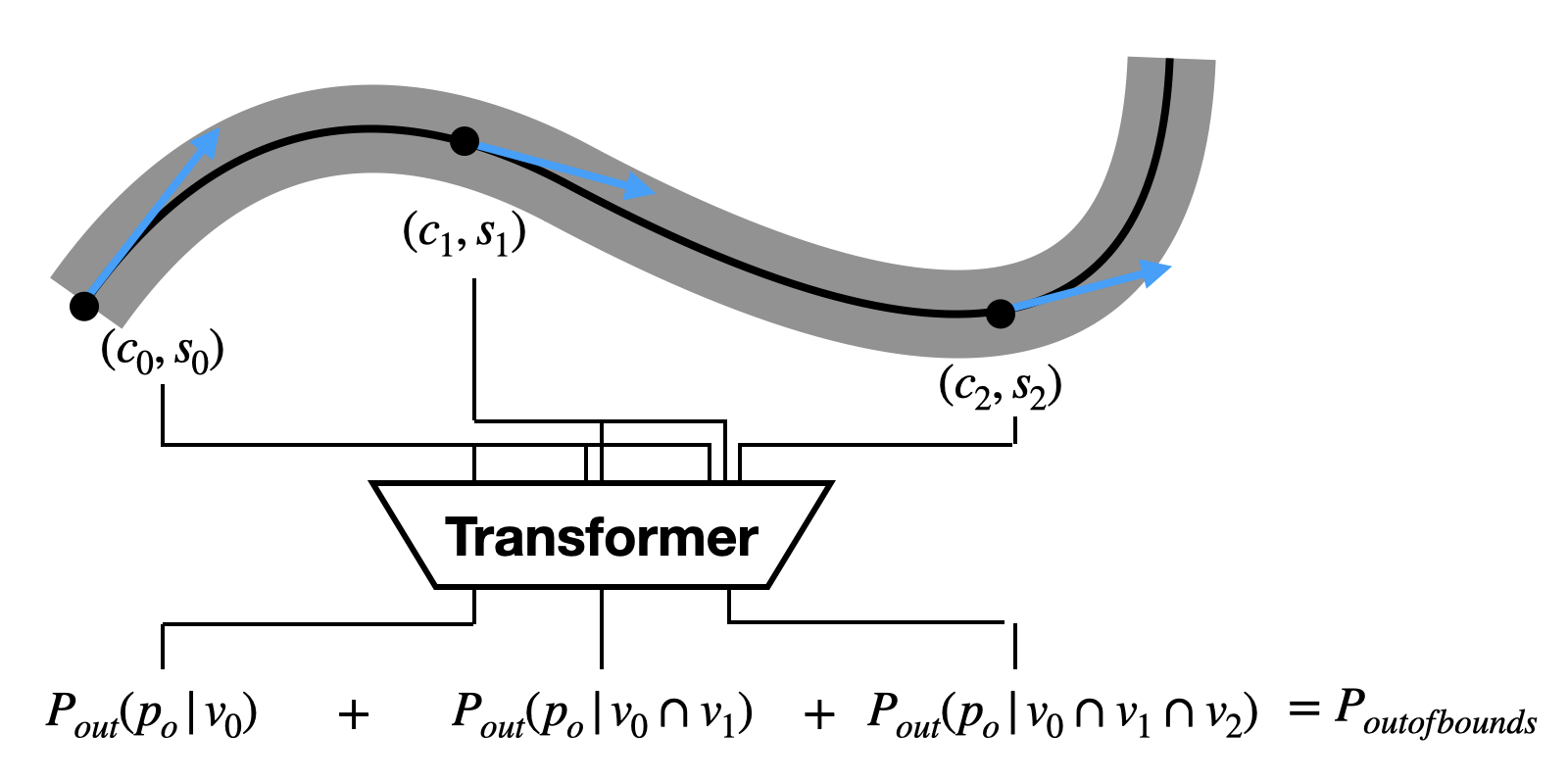}
    \caption{The discriminator model takes the previous road points (curvature, step size) as input to predict for each point the likelihood of an OOB condition}
    \label{fig:transformer}
    \vspace{-5mm}
\end{figure}

We use the summed probability output of being out of bound as a fitness function $F_1$ for each road (Figure~\ref{fig:transformer}), which we use in our genetic algorithm.
As a second fitness metric $F_2$, we use the median Euclidian distance to every other road segment in the pool, which represents our diversity metric.
We trained the discriminator for 500 epochs until we reached our peak sensitivity of $0.52$ and specificity of $0.96$ in our validation set. 

\subsection{Genetic Algorithm} \label{Genetic_Algorithm}

For our genetic algorithm, we start with AmbieGen as a template but adapt it to our problem.
We encode our road as a series of 50 curvature points $c_i$ with length cumulative length $s_i$. Therefore, we use the following operators:

\begin{enumerate}
    \item Crossover operator: Given two roads, a curvature index is randomly chosen, splitting the roads in half with an exchange of segments;
    \item K-Crossover operator: It operates as the crossover operator but with two splitting points;
    \item Swap operator: We take two random segments of the road between lengths 5 and 15 and interchange them;
    \item Mutation operator: We randomly chose an index, sample a random curvature value between $[-0.7, 0.7]$, and set the index $+-3$ to the curvature value. This has the effect of randomly introducing a curve to the road.
\end{enumerate}

We initialize a population of random roads where, at each iteration, there is a probability of 0.8 for the crossover operator, 0.4 for 2-crossover, 0.4 for swap, and 0.2 for mutation.
We then select the top 3,000 roads based on our $F_1$ fitness score and further select 2,000 of our population based on our diversity metric $F_2$.
This has the effect of prioritizing fitness but maintaining diversity.

We observed that roads with high fitness were characterized by non-smooth segments.
Consequently, we added a univariate cubic spline interpolation (smoothing factor: 0.01) for the newly generated road segments to produce naturally smooth road segments.
To avoid duplicates, we also check for matching road segments that have an Euclidean distance of less than 0.2, which we consequently exclude from reproduction.
A secondary validity test checks for curvature constraints $[-0.1; 0.1]$, boundary constraints (inside of the 200x200 limit), and not self-intersecting, which are consequently removed from the pool.
We run the algorithm for $50$ epochs, reaching a mean probability of being out of bounds of $33\%$ and a median distance of $0.4$ to every other road segment.

\section{Empirical evaluation}  
To empirically evaluate \approach, we apply the same experimental setting of \frenetic at the SBST tool competition~\cite{PanichellaGZR21} to compare the results.

\subsection{Research question}
In the first part of the evaluation of \approach, we want to know to what extent the approach generates feasible test cases.
The \beamng simulator and the test runner have certain restrictions, e.g., the test case must not contain roads that intersect with themselves or that comprise curves being too sharp.
If the test does not fulfill the abovementioned requirements, then it is considered an \textit{invalid test case}.
Thus, our first research question is:

\vspace*{5px}
\begin{mdframed}
\RQ{1}: \textit{Does diversity-guided search exploration in Frenet space with \approach lead to more valid test cases compared to Frenetic?}
\end{mdframed}
\vspace*{5px}

We investigate to what extent our approach produces valid test cases compared to Frenetic.
An approach that produces many invalid test cases is not optimal, especially considering that the generation process requires a significant amount of time/testing budget.
Assuming the approach produces only a few invalid test cases, we can focus on effective testing, i.e., test cases that will eventually fail.
Hence, our second research question is as follows:

\vspace*{5px}
\begin{mdframed}
\RQ{2}: \textit{Does diversity-guided search exploration in Frenet space with \approach lead to higher fault detection compared to Frenetic?}
\end{mdframed}
\vspace*{5px}

We execute 1,174 test cases generated by \approach.
By measuring the execution time and logging the test outcomes, we compute the number of test failures, i.e., the approach's effectiveness. 

\subsection{Test subject}
We use the \beamng simulation environment with its built-in self-driving agent to evaluate \approach.
Concretely, we use the \beamng \textit{v0.24.0.2}, which is a more recent version than the version in the context of SBST'21.
The driving agent steers the car with the goal of keeping the SDC within the lane.
The SDC must stay within the lane by at least 85\% of its area; otherwise, the test will fail.

\subsection{Procedure}
To address \RQ{1} and \RQ{2}, we compare the results of \frenetic with \approach, and apply the same procedure as in SBST'21~\cite{DBLP:conf/sbst/2021}.
We provide \approach to generate and execute test cases with a time budget of two hours.
The driving agent is configured to have a maximum speed of 70km/h, and the oracle was set to an out-of-bound (OOB)~\cite{10.1145/3533818,DBLP:journals/ese/BirchlerKBGP23} of 30\%.

\textbf{Test generation.}
We generated 1,174 test cases with our approach \approach.
The tests are stored as JSON files compatible with the \scissor tool to execute the tests in \beamng.

\textbf{Test execution.}
We execute all 1,174 generated test cases in \beamng.
For configuring the runtime parameters, such as the OOB tolerance and test runner, we use \scissor~\cite{9825849}, which is a reimplementation of the test runner provided by SBST'21 with more up-to-date dependencies.

\textbf{Infrastructure.}
We used a Windows 10 machine with an Intel(R) Core(TM) i9-10900KF CPU at 3.70GHz, 32 GB RAM, and a 10GB NVIDIA GeForce RTX 3080 graphics card.

\subsection{Data collection}
For each test execution, we log the test outcome and simulation time in JSON files.
Concretely, we have in the JSON file the following fields:
\begin{inparaenum}[(i)]
    \item \texttt{test\_outcome}, which can be either ``PASS'' or ``FAIL'', and 
    \item \texttt{test\_duration}, for the simulation time measured in seconds.
\end{inparaenum}

\subsection{Data analysis}
We count the number of faults detected during the test execution as well the number of invalid test cases produced by \approach within the given time budget.
Since we ran 1,174 test cases, which require more execution time than the time budget limit, we randomly selected the test results with the logged simulation time until we reached the time budget of two hours.
We performed the random selection with replacement 100 times and computed statistics.
Concretely, we compare the number of invalid test cases (\RQ{1}) and fault detection rate (\RQ{2}) with the results of \frenetic from~\cite{PanichellaGZR21}.

All data and code are available in a replication package (Section~\ref{sec:data-availability}).
For the analysis, we set up a \textsc{MongoDB} with the simulations data to effectively query the data with our analysis scripts.

\begin{table}[b]
    \centering
    \caption{Descriptive statistics of 100 samples with a simulation time budget of two hours.}
    \begin{tabular}{l|lll} \toprule
    \textbf{Statistic} & \textbf{\# Executed} & \textbf{\# Invalid} & \textbf{\# Faults} \\ \midrule
    \textit{Min.} & 778      & 0 & 482 \\
    \textit{Avg.} & 798.47   & 0 & 508.58 \\
    \textit{Max.} & 815      & 0 & 536 \\ \bottomrule
    \end{tabular}
    \label{tab:results}
\end{table}

\section{Results \& Discussion}
With \approach, we ran a total of 1,174 test cases, which is also our population of test cases from which we randomly sample sets of test cases, where each set consists of test cases with a total simulation time within the time budget, i.e., we construct a sample set by taking tests from the population until we reach the time budget.
We sampled 100 times with replacement so that we can compute the statistics, such as the minimum, maximum, and mean of invalid and failing test cases.
As indicated in Table~\ref{tab:results}, from all 100 samples, we did not identify any invalid road, i.e., a road that intersects with itself.
Considering large testing campaigns, the \approach is able to generate valid test cases effectively and does not introduce overhead in the generation process by generating invalid test cases.
\approach produces failing tests in most cases;
on average, \approach generates 508.58 test cases leading to an OOB violation within a time budget of two hours.
When taking into consideration that, within the budget, 796.47 tests are executed on average, we have 64\% failing test cases.
In the tool competition~\cite{PanichellaGZR21}, \frenetic generated on average 20-25 (boxplot in \cite{PanichellaGZR21}) failing test cases.
With \approach, we have an approach that effectively generates test cases, which helps to reduce the testing costs since we spend less time on executing passing (``uninformative'') test cases.

\subsection{Threats to validity}

Due to the large number of training examples in comparison to the number of tests we generate, we find a number of generated tests that appear similar to the ones in our training set.
We derived the mean minimal distance for every test to every training example, which we find to be an average of $0.15$, which is below our duplication criterium of $0.2$.
However, the average median distance (our diversity measure) is higher, i.e., at $0.6$, showing a relatively high diversity.
It is important to note that there is no explicit conditioning of our discriminator model on the generation of tests, but that through the fitness metric, the model might favor examples that appear similar to what it has seen in its training.
Nevertheless, we also find $300$ examples with a minimal distance above $0.2$, proving that we are able to generate completely novel tests.

Regarding \RQ{1}, an invalid road is defined as a road that intersects with itself or has overly sharp turns.
As far as we know, no well-defined definition of an overly sharp turn in the literature is available.
Hence, we can not exclude that \approach would still generate invalid roads.
Furthermore, with respect to \RQ{2}, \approach would benefit from an evaluation with statistical tests to show its performance against the benchmark with more confidence. 
We did not perform experiments with \frenetic but only used the reported results from \cite{PanichellaGZR21}.
Thus, the comparison of \approach with \frenetic would benefit from a very identical experimental setup.

\section{Conclusion}
In summary, we have shown that the OOB metric can be implicitly learned by a model without explicitly formulating the problem.
In combination with a genetic algorithm, we have shown that a diverse population of tests can be generated, showing high validity and relatively high fault-revealing power.

Future works should focus on leveraging the ability of transformer models to work with variable road lengths, possibly increasing the diversity of tests that can be generated.
We could improve on diversity by extending the diversity measure to the training set and generated test, thus favoring more diversity overall.

Currently, we also rely on a large number of failing tests, which are needed beforehand.
A reinforcement model that continuously improves its ability to predict the likelihood of failure could work as a discriminator for a genetic algorithm, identifying the tests that might produce failures before having to execute them. 


\section{Data availability}\label{sec:data-availability}
The code and dataset are publicly available on Zenodo~\cite{blattner_2024_10527296}.

\section*{CRediT author statement} 
\textbf{Timo Blattner}: Conceptualization, Methodology, Software, Validation, Formal analysis, Investigation, Writing - Original Draft.
\textbf{Christian Birchler}: Investigation, Resources, Data Curation, Writing - Original Draft.
\textbf{Timo Kehrer}: Writing - Review \& Editing, Funding acquisition.
\textbf{Sebastiano Panichella}: Writing Original Draft, Supervision, Funding acquisition.

\begin{acks}
We thank the Horizon 2020 (EU Commission) support for the \href{https://www.cosmos-devops.org/}{project COSMOS}, 
Project No. 957254-COSMOS.
\end{acks}

\bibliographystyle{ACM-Reference-Format}
\bibliography{main}


\begin{thebibliography}{17}


\ifx \showCODEN    \undefined \def \showCODEN     #1{\unskip}     \fi
\ifx \showDOI      \undefined \def \showDOI       #1{#1}\fi
\ifx \showISBNx    \undefined \def \showISBNx     #1{\unskip}     \fi
\ifx \showISBNxiii \undefined \def \showISBNxiii  #1{\unskip}     \fi
\ifx \showISSN     \undefined \def \showISSN      #1{\unskip}     \fi
\ifx \showLCCN     \undefined \def \showLCCN      #1{\unskip}     \fi
\ifx \shownote     \undefined \def \shownote      #1{#1}          \fi
\ifx \showarticletitle \undefined \def \showarticletitle #1{#1}   \fi
\ifx \showURL      \undefined \def \showURL       {\relax}        \fi
\providecommand\bibfield[2]{#2}
\providecommand\bibinfo[2]{#2}
\providecommand\natexlab[1]{#1}
\providecommand\showeprint[2][]{arXiv:#2}

\bibitem[DBL(2021)]%
        {DBLP:conf/sbst/2021}
 \bibinfo{year}{2021}\natexlab{}.
\newblock \bibinfo{booktitle}{\emph{Intl. Workshop on Search-Based Software
  Testing}}. \bibinfo{publisher}{{IEEE}}.
\newblock
\showISBNx{978-1-6654-4571-9}
\urldef\tempurl%
\url{https://doi.org/10.1109/SBST52555.2021}
\showDOI{\tempurl}


\bibitem[Administration(2022)]%
        {us-adas-accidents}
\bibfield{author}{\bibinfo{person}{National Highway Traffic~Safety
  Administration}.} \bibinfo{year}{2022}\natexlab{}.
\newblock \bibinfo{booktitle}{\emph{Summary Report: Standing General Order on
  Crash Reporting for Level 2 Advanced Driver Assistance Systems}}.
\newblock \bibinfo{type}{{T}echnical {R}eport} DOT HS 813 325.
  \bibinfo{institution}{National Highway Traffic Safety Administration},
  \bibinfo{address}{1200 New Jersey Avenue, SE Washington, D.C. 20590}.
\newblock


\bibitem[Biagiola et~al\mbox{.}(2023)]%
        {10190377}
\bibfield{author}{\bibinfo{person}{Matteo Biagiola}, \bibinfo{person}{Stefan
  Klikovits}, \bibinfo{person}{Jarkko Peltomäki}, {and}
  \bibinfo{person}{Vincenzo Riccio}.} \bibinfo{year}{2023}\natexlab{}.
\newblock \showarticletitle{SBFT Tool Competition 2023 - Cyber-Physical Systems
  Track}. In \bibinfo{booktitle}{\emph{Intl. Workshop on Search-Based and Fuzz
  Testing (SBFT)}}. \bibinfo{pages}{45--48}.
\newblock
\urldef\tempurl%
\url{https://doi.org/10.1109/SBFT59156.2023.00010}
\showDOI{\tempurl}


\bibitem[Birchler et~al\mbox{.}(2022)]%
        {9825849}
\bibfield{author}{\bibinfo{person}{Christian Birchler},
  \bibinfo{person}{Nicolas Ganz}, \bibinfo{person}{Sajad Khatiri},
  \bibinfo{person}{Alessio Gambi}, {and} \bibinfo{person}{Sebastiano
  Panichella}.} \bibinfo{year}{2022}\natexlab{}.
\newblock \showarticletitle{Cost-effective Simulation-based Test Selection in
  Self-driving Cars Software with SDC-Scissor}. In
  \bibinfo{booktitle}{\emph{Intl. Conference on Software Analysis, Evolution
  and Reengineering}}. \bibinfo{publisher}{{IEEE}}, \bibinfo{pages}{164--168}.
\newblock
\urldef\tempurl%
\url{https://doi.org/10.1109/SANER53432.2022.00030}
\showDOI{\tempurl}


\bibitem[Birchler et~al\mbox{.}(2023a)]%
        {DBLP:journals/ese/BirchlerKBGP23}
\bibfield{author}{\bibinfo{person}{Christian Birchler}, \bibinfo{person}{Sajad
  Khatiri}, \bibinfo{person}{Bill Bosshard}, \bibinfo{person}{Alessio Gambi},
  {and} \bibinfo{person}{Sebastiano Panichella}.}
  \bibinfo{year}{2023}\natexlab{a}.
\newblock \showarticletitle{Machine learning-based test selection for
  simulation-based testing of self-driving cars software}.
\newblock \bibinfo{journal}{\emph{Empir. Softw. Eng.}} \bibinfo{volume}{28},
  \bibinfo{number}{3} (\bibinfo{year}{2023}), \bibinfo{pages}{71}.
\newblock
\urldef\tempurl%
\url{https://doi.org/10.1007/S10664-023-10286-Y}
\showDOI{\tempurl}


\bibitem[Birchler et~al\mbox{.}(2023b)]%
        {10.1145/3533818}
\bibfield{author}{\bibinfo{person}{Christian Birchler}, \bibinfo{person}{Sajad
  Khatiri}, \bibinfo{person}{Pouria Derakhshanfar}, \bibinfo{person}{Sebastiano
  Panichella}, {and} \bibinfo{person}{Annibale Panichella}.}
  \bibinfo{year}{2023}\natexlab{b}.
\newblock \showarticletitle{Single and Multi-Objective Test Cases
  Prioritization for Self-Driving Cars in Virtual Environments}.
\newblock \bibinfo{journal}{\emph{ACM Trans. Softw. Eng. Methodol.}}
  \bibinfo{volume}{32}, \bibinfo{number}{2} (\bibinfo{date}{apr}
  \bibinfo{year}{2023}).
\newblock
\showISSN{1049-331X}
\urldef\tempurl%
\url{https://doi.org/10.1145/3533818}
\showDOI{\tempurl}


\bibitem[Blattner et~al\mbox{.}(2024)]%
        {blattner_2024_10527296}
\bibfield{author}{\bibinfo{person}{Timo Blattner}, \bibinfo{person}{Christian
  Birchler}, \bibinfo{person}{Timo Kehrer}, {and} \bibinfo{person}{Sebastiano
  Panichella}.} \bibinfo{year}{2024}\natexlab{}.
\newblock \bibinfo{title}{{Diversity-guided Search Exploration for Self-
  driving Cars Test Generation through Frenet Space Encoding [REPLICATION
  PACKAGE]}}.
\newblock
\newblock
\urldef\tempurl%
\url{https://doi.org/10.5281/zenodo.10527295}
\showDOI{\tempurl}


\bibitem[Castellano et~al\mbox{.}(2021)]%
        {DBLP:conf/sbst/CastellanoCTKZA21}
\bibfield{author}{\bibinfo{person}{Ezequiel Castellano}, \bibinfo{person}{Ahmet
  Cetinkaya}, \bibinfo{person}{C{\'{e}}dric~Ho Thanh}, \bibinfo{person}{Stefan
  Klikovits}, \bibinfo{person}{Xiaoyi Zhang}, {and} \bibinfo{person}{Paolo
  Arcaini}.} \bibinfo{year}{2021}\natexlab{}.
\newblock \showarticletitle{Frenetic at the {SBST} 2021 Tool Competition}. In
  \bibinfo{booktitle}{\emph{Intl. Workshop on Search-Based Software Testing,
  {SBST} 2021, Madrid, Spain, May 31, 2021}}. \bibinfo{publisher}{{IEEE}},
  \bibinfo{pages}{36--37}.
\newblock
\urldef\tempurl%
\url{https://doi.org/10.1109/SBST52555.2021.00016}
\showDOI{\tempurl}


\bibitem[Di~Sorbo et~al\mbox{.}(2023)]%
        {UAVtosem}
\bibfield{author}{\bibinfo{person}{Andrea Di~Sorbo}, \bibinfo{person}{Fiorella
  Zampetti}, \bibinfo{person}{Aaron Visaggio}, \bibinfo{person}{Massimiliano
  Di~Penta}, {and} \bibinfo{person}{Sebastiano Panichella}.}
  \bibinfo{year}{2023}\natexlab{}.
\newblock \showarticletitle{Automated Identification and Qualitative
  Characterization of Safety Concerns Reported in UAV Software Platforms}.
\newblock \bibinfo{journal}{\emph{ACM Trans. Softw. Eng. Methodol.}}
  \bibinfo{volume}{32}, \bibinfo{number}{3} (\bibinfo{year}{2023}).
\newblock
\showISSN{1049-331X}
\urldef\tempurl%
\url{https://doi.org/10.1145/3564821}
\showDOI{\tempurl}


\bibitem[Khatiri et~al\mbox{.}(2023a)]%
        {khatiri2023simulation}
\bibfield{author}{\bibinfo{person}{Sajad Khatiri}, \bibinfo{person}{Sebastiano
  Panichella}, {and} \bibinfo{person}{Paolo Tonella}.}
  \bibinfo{year}{2023}\natexlab{a}.
\newblock \showarticletitle{Simulation-based test case generation for unmanned
  aerial vehicles in the neighborhood of real flights}. In
  \bibinfo{booktitle}{\emph{Intl. Conference on Software Testing, Verification
  and Validation (ICST)}}.
\newblock


\bibitem[Khatiri et~al\mbox{.}(2023b)]%
        {KhatiriPT23}
\bibfield{author}{\bibinfo{person}{Sajad Khatiri}, \bibinfo{person}{Sebastiano
  Panichella}, {and} \bibinfo{person}{Paolo Tonella}.}
  \bibinfo{year}{2023}\natexlab{b}.
\newblock \showarticletitle{Simulation-based Test Case Generation for Unmanned
  Aerial Vehicles in the Neighborhood of Real Flights}. In
  \bibinfo{booktitle}{\emph{Conference on Software Testing, Verification and
  Validation}}. \bibinfo{publisher}{{IEEE}}, \bibinfo{pages}{281--292}.
\newblock
\urldef\tempurl%
\url{https://doi.org/10.1109/ICST57152.2023.00034}
\showDOI{\tempurl}


\bibitem[Khatiri et~al\mbox{.}(2024)]%
        {SBFT-UAV2024}
\bibfield{author}{\bibinfo{person}{Sajad Khatiri}, \bibinfo{person}{Prasun
  Saurabh}, \bibinfo{person}{Timothy Zimmermann}, \bibinfo{person}{Charith
  Munasinghe}, \bibinfo{person}{Christian Birchler}, {and}
  \bibinfo{person}{Sebastiano Panichella}.} \bibinfo{year}{2024}\natexlab{}.
\newblock \showarticletitle{{SBFT} Tool Competition 2024 - CPS-UAV Test Case
  Generation Track}. In \bibinfo{booktitle}{\emph{Intl. Workshop on
  Search-Based and Fuzz Testing, SBFT@ICSE 2024}}.
\newblock


\bibitem[Klikovits et~al\mbox{.}(2023)]%
        {KlikovitsCCA23}
\bibfield{author}{\bibinfo{person}{Stefan Klikovits}, \bibinfo{person}{Ezequiel
  Castellano}, \bibinfo{person}{Ahmet Cetinkaya}, {and} \bibinfo{person}{Paolo
  Arcaini}.} \bibinfo{year}{2023}\natexlab{}.
\newblock \showarticletitle{Frenetic-lib: An extensible framework for
  search-based generation of road structures for {ADS} testing}.
\newblock \bibinfo{journal}{\emph{Sci. Comput. Program.}}
  \bibinfo{volume}{230} (\bibinfo{year}{2023}), \bibinfo{pages}{102996}.
\newblock
\urldef\tempurl%
\url{https://doi.org/10.1016/J.SCICO.2023.102996}
\showDOI{\tempurl}


\bibitem[{NPR}({[n.\,d.]})]%
        {NPR}
\bibfield{author}{\bibinfo{person}{{NPR}}.}
  \bibinfo{year}{[n.\,d.]}\natexlab{}.
\newblock \bibinfo{title}{Nearly 400 car crashes in 11 months involved
  automated tech, companies tell regulators}.
\newblock \bibinfo{howpublished}{npr.org}.
\newblock
\urldef\tempurl%
\url{https://www.npr.org/2022/06/15/1105252793/nearly-400-car-
  crashes-in-11-months-involved-automated-tech-companies-tell- regul}
\showURL{%
\tempurl}


\bibitem[Panichella et~al\mbox{.}(2021)]%
        {PanichellaGZR21}
\bibfield{author}{\bibinfo{person}{Sebastiano Panichella},
  \bibinfo{person}{Alessio Gambi}, \bibinfo{person}{Fiorella Zampetti}, {and}
  \bibinfo{person}{Vincenzo Riccio}.} \bibinfo{year}{2021}\natexlab{}.
\newblock \showarticletitle{{SBST} Tool Competition 2021}. In
  \bibinfo{booktitle}{\emph{Intl. Workshop on Search-Based Software Testing,
  {SBST} 2021, Madrid, Spain, May 31, 2021}}. \bibinfo{publisher}{{IEEE}},
  \bibinfo{pages}{20--27}.
\newblock
\urldef\tempurl%
\url{https://doi.org/10.1109/SBST52555.2021.00011}
\showDOI{\tempurl}


\bibitem[Peltom{\"a}ki et~al\mbox{.}(2022)]%
        {peltomaki2022wasserstein}
\bibfield{author}{\bibinfo{person}{Jarkko Peltom{\"a}ki},
  \bibinfo{person}{Frankie Spencer}, {and} \bibinfo{person}{Ivan Porres}.}
  \bibinfo{year}{2022}\natexlab{}.
\newblock \showarticletitle{Wasserstein generative adversarial networks for
  online test generation for cyber physical systems}. In
  \bibinfo{booktitle}{\emph{Workshop on Search-Based Software Testing}}.
  \bibinfo{pages}{1--5}.
\newblock


\bibitem[Winsten and Porres(2023)]%
        {WinstenP23}
\bibfield{author}{\bibinfo{person}{Jesper Winsten} {and} \bibinfo{person}{Ivan
  Porres}.} \bibinfo{year}{2023}\natexlab{}.
\newblock \showarticletitle{{WOGAN} at the {SBFT} 2023 Tool Competition -
  Cyber-Physical Systems Track}. In \bibinfo{booktitle}{\emph{Intl. Workshop on
  Search-Based and Fuzz Testing, SBFT@ICSE 2023, Melbourne, Australia, May 14,
  2023}}. \bibinfo{publisher}{{IEEE}}, \bibinfo{pages}{43--44}.
\newblock
\urldef\tempurl%
\url{https://doi.org/10.1109/SBFT59156.2023.00009}
\showDOI{\tempurl}


\end{thebibliography}

\end{document}